\documentclass[12pt]{iopart}
\usepackage{graphicx}
\usepackage{color}

\usepackage{color}
\definecolor{dblue}{rgb}{0,0,0.75}
\definecolor{dred}{rgb}{0.6,0,0}
\definecolor{dgreen}{rgb}{0,0.5,0}

\begin{document}

\title{Nonlinear q-voter model with deadlocks on the Watts-Strogatz graph}
\author{Katarzyna Sznajd-Weron$^1$, Karol Michal Suszczynski $^{2}$}
\address{$^1$ Institute of Physics, Wroc{\l}aw University of Technology, Poland\\ 
$^2$ Institute of Theoretical Physics, University of Wroc{\l}aw, Poland}
\begin{abstract}
We study the nonlinear $q$-voter model with deadlocks on a Watts-Strogats graph. Using Monte Carlo simulations, we obtain so called exit probability and exit time. We determine how network properties, such as randomness or density of links influence exit properties of a model. 
\end{abstract}

\section{Introduction}
Describing opinion dynamics has inspired many physicists to build models that could not be justified by physical phenomena (for review of opinion dynamic models see \cite{Cas:For:Lor:09,Gal:12}). Such models are usually rather caricatures than precise portraits of real social systems. However, far going simplifications should not be regarded as a defect of these models. Simplicity allows not only for in-depth analysis or sometimes even analytical treatment. First of all it allows to describe some universal features or even to determine the most important factors that influence given social phenomenon. 

Certainly, the main challenge that stays behind opinion dynamics models is to describe complex social systems in terms of a relatively simple approach. On the other hand, such models are itself interesting from a theoretical point of view \cite{Kra:Red:Ben:10}. Therefore they might be also treated as small building blocks which give a contribution to the construction of still emerging non-equilibrium statistical physics. One of a good example of such an interesting model is nonlinear $q$-voter model introduced by \cite{Cas:Mun:Pas:09} and its modifications proposed by \cite{Prz:Szn:Tab:11,Nyc:Szn:Cis12}. In this paper we will investigate a special case of a model, which we call $q$-voter model with deadlocks, considered already by \cite{Prz:Szn:Tab:11} in a case of a one dimensional lattice. In this paper we examine the role of a topology in such a model and show that increasing randomness and density of a network helps to reach consensus.

\section{Model}
Original $q$-voter model introduced in \cite{Cas:Mun:Pas:09} has been defined as follows:
\begin{itemize}
\item Each $i$-th site of a graph of a size $N$ is occupied by a voter $S_i = \pm 1$
\item Initially there is a probability $p$ of finding a voter in a state $+1$ and probability $1-p$ finding a voter in a state $-1$
\item System evolves according to the following algorithm:
\begin{enumerate}
\item At each elementary time step $t$, choose one spin $S_i$, located at site $i$, at random
\item Choose $q$ neighbors ($q$-panel) of site $i$
\item If all $q$ neighbors are in the same state than $S_i$ takes the value of the $q$ neighbors
\item Otherwise, if the $q$ neighbors are not unanimous than $S_i \rightarrow -S_i$ with probability $\epsilon$
\item Time is updated $t \rightarrow t+\frac{1}{N}$
\end{enumerate} 
\end{itemize}
In \cite{Prz:Szn:Tab:11} it has been proposed to study a one-dimensional model for $\epsilon=0$, which for $q=2$ corresponds to the Sznajd model, as noted in \cite{Cas:Mun:Pas:09}. It should be noted that for $\epsilon=0$ the evolution of the system is hampered due to the existence of deadlock configurations. Deadlocks should be understood as configurations in which there is no possibility for an evolution due to the lack of an unanimous $q$-panel. In a case of a one dimensional lattice and $q=2$ there is only one deadlock configuration -- antiferromagnetic state $+-+-+...$. For $q=3$ there are already much more e.g. $+-+-+-,...$, $++--++--++...$ or $++-+-++-+...$, etc. However, if initially there is at least one $q$-panel, evolution will reach one of two final absorbing states. The nonlinear $q$-voter model with deadlocks has been found to be interesting for several reasons:
\begin{itemize}
\item For $q=2$ it reduces to Sznajd model for opinion dynamics and in this case the analytical formula for an exit probability has been found independently by \cite{Lam:Red:2008,Sla:Szn:Prz:08,Cas:Pas:11,Gal:Mar:11}
\item Exit probability does not depend on a system size as reported by \cite{Prz:Szn:Tab:11}. This result has to treated cautiously taking into account recent results obtained by \cite{Tim:Pra:13} for large lattices. It seems that additional simulations are needed to explain this contradiction. 
\item In a case of a random noise, system undergoes a phase transition which changes its type from continuous to discontinuous at $q=5$ \cite{Nyc:Szn:Cis12}.  
\end{itemize} 
To investigate the role of the network topology we have decided to use the model introduced by \cite{Wat:Str:98}, mainly because it allows to study various structures -- from regular lattices with different size of the neighborhood, through the small-world networks to random graphs. The Watts-Strogatz algorithm, that we have used, is defined as follows:
\begin{itemize}
\item Start with a 1D lattice of size $N$ with periodic boundary conditions in which each node is connected to its $k$ neighbors 
\item Then with probability $\beta$ replace each edge by a randomly chosen edge
\end{itemize}
For $\beta=0$ we deal with regular lattices -- e.g. for $k=1$ we have simple one-dimensional lattice with integrations only to the nearest neighbors and for $k=N/2-1$ we have a complete graph. With increasing $\beta$ we increase randomness of the network going through the small-world (for $\beta=0.01 - 0.1$) to random graph for $\beta=1$. Summarizing, we have one parameter $q$ that defines the model itself and two parameters $k$ and $\beta$ that describe the network properties. 

\section{Results}
\subsection{Exit probability on a complete graph}
One of the important properties of the system with the absorbing states is so called exit probability \cite{Kra:Red:Ben:10}. In our case exit probability $E(p)$ should be understood as a probability of the absorbing state with all spins '+1' as a function of the initial probability $p$ of finding a spin in a state $+1$.
For a complete graph, which corresponds also to the mean-field treatment, the evolution of the probability of 'up'-spins is given by:
\begin{equation}
p(t+\Delta t)=p(t)+p^q(t)(1-p(t))-(1-p(t))^qp(t)
\label{mfa_evol}
\end{equation}
Fixed points can be easily found from the condition $p(t+\Delta t)=p(t)=p^*$, i.e.:
\begin{equation}
(p^*)^q(1-p^*)-(1-p^*)^qp^*=p^*(1-p^*)\left[ (p^*)^{q-1}-(1-p^*)^{q-1} \right]=0.
\label{mfa_evol1}
\end{equation}
As can be seen there are three fixed points $p*=0,1/2,1$. It can be easily check calculating the following derivative:
\begin{equation}
\frac{d}{dp}\left( p+p^q(1-p)-(1-p)^qp \right)|_{p=p^*},
\end{equation}
that $p^*=0$ and $p^*=1$ are stable, whereas $p^*=1/2$ is unstable fixed point. Therefore on a complete graph for $p<1/2$ the system eventually reaches absorbing state $p^*=0$ and for $p>1/2$ system reaches $p^*=1$. This means that the exit probability for the $q$-voter model on the complete graph with arbitrary value of $q$ is a step-like function:
\begin{equation}
E(p)=\left\{
\begin{array}{ccc}
0 & \mbox{for} & p<1/2\\
1 & \mbox{for} & p>1/2\\
\end{array}
\right.
\end{equation}
\cite{Tim:Pra:13} have recently proposed much more rigorous approach to show that the exit probability is a step-like function for a complete graph. Their and our results confirm that the results obtained by \cite{Gal:05} within his unifying frame (GUF) coincides with the mean-field approach and may not be true for arbitrary topology.

\subsection{Results on a Watts-Strogatz graph}
It has been shown independently by \cite{Lam:Red:2008,Sla:Szn:Prz:08,Cas:Pas:11,Gal:Mar:11} that for $q=2$ on a one-dimensional lattice (which corresponds to the Sznajd model) exit probability is given by:
\begin{equation}
E(p)=\frac{p^2}{p^2+(1-p)^2}.
\label{Ep2}
\end{equation}
In \cite{Prz:Szn:Tab:11} this result has been generalized for arbitrary value of $q$ in a case of a one-dimensional lattice ($k=1, \beta=0$):
\begin{equation}
E(p)=\frac{p^q}{p^q+(1-p)^q}.
\label{eq:Epq}
\end{equation}
Here we would like to examine the role of the network properties given by parameters $k$ and $\beta$. We start with $q=2$ and $\beta=0$, i.e. regular lattices with various size of the neighborhood (i.e. various $k$).

\begin{figure}
\begin{center}
\includegraphics[width=14cm]{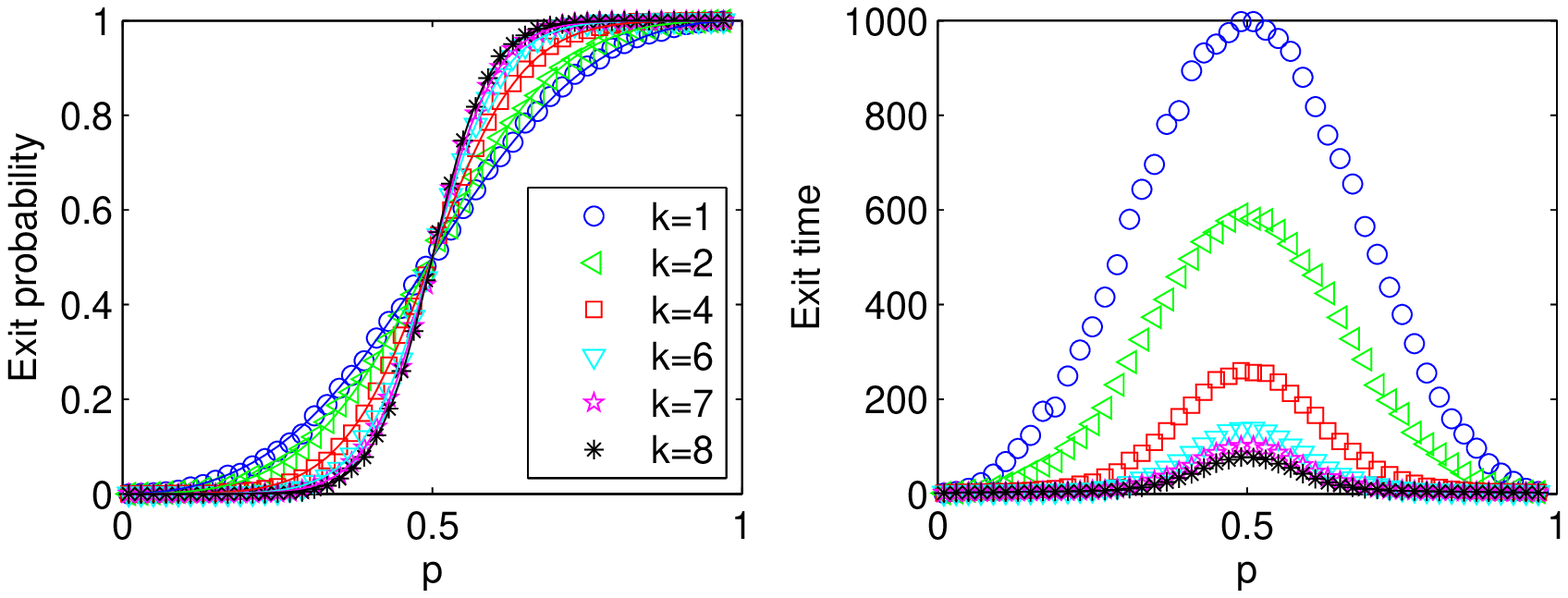}
\end{center}
 \textbf{\refstepcounter{figure}\label{fig:01} Figure \arabic{figure}.}{Exit probability (left panel) and exit time (right panel) as a function of the initial probability $p$ of spin $+1$ for $q=2$ (which corresponds to the Sznajd model) and $\beta=0$ (regular graph) for several sizes of the neighborhood given by $k$. The system size $N=100$ and results were averaged over $10^4$ samples. Solid lines on the left panel correspond to analytical formula given by Eq. (\ref{eq:anal_f}). The steepness of the exit probability slope increases with $k$ and exit time significantly decreases with $k$.}
\end{figure}

It is seen (see Fig.1) that the steepness of the exit probability slope increases with $k$, which is expected because we have found that for $k=N/2-1$ (complete graph) $E(p)$ is a step-like function. Moreover, it occurs that for arbitrary value of $k$ simulation data can be fitted by:
\begin{equation}
E(p)=\frac{p^f}{p^f+(1-p)^f},
\label{eq:anal_f}
\end{equation}
where $f=k/2+3/2$.
Parameter $k$ influences also exit time, which should be understood as a time needed to reach an absorbing state \cite{ Kra:Red:Ben:10}. As seen the exit time significantly decreases with $k$, which means that consensus is reached faster for more dense societies.

\begin{figure}
\begin{center}
\includegraphics[width=14cm]{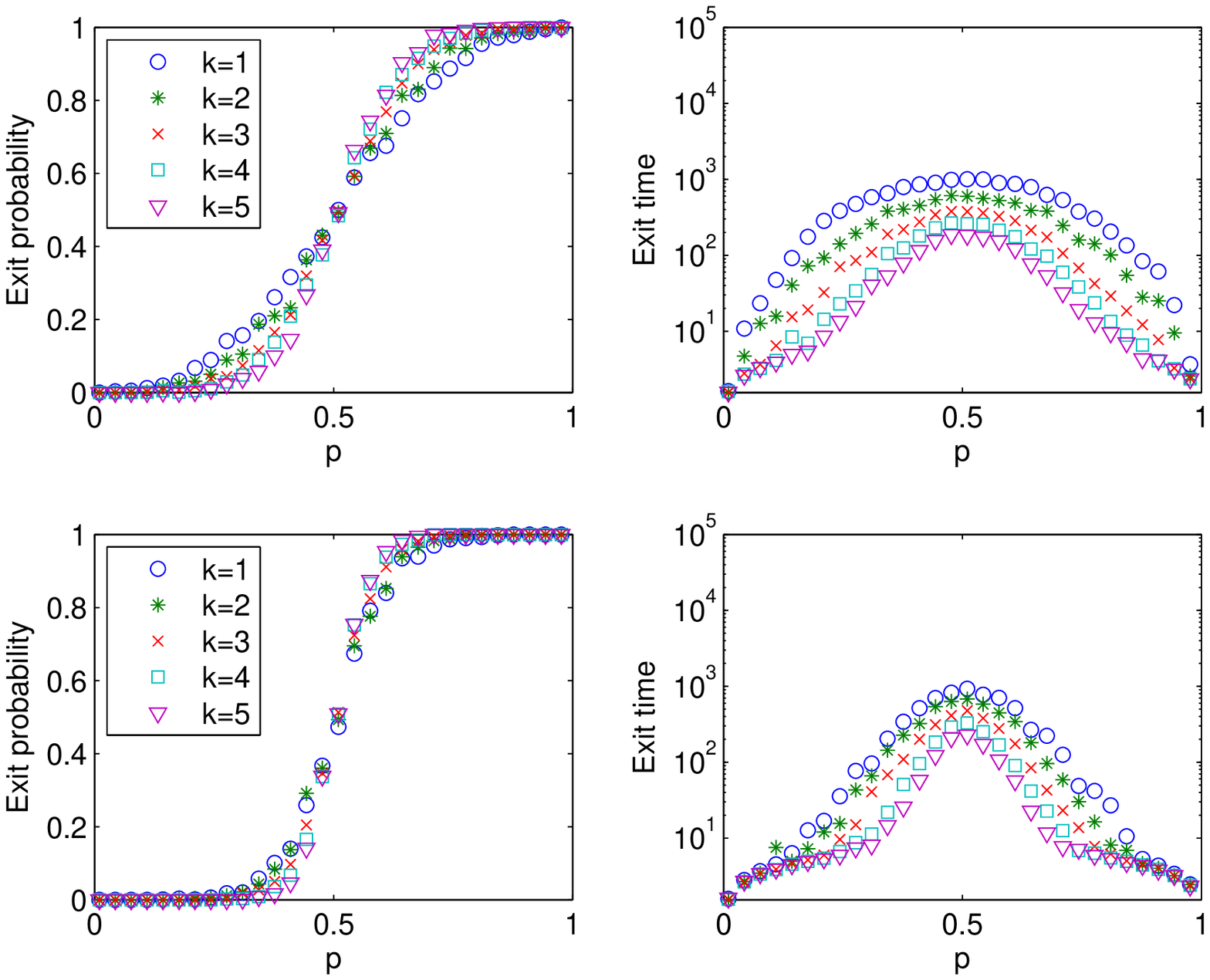}
\end{center}
 \textbf{\refstepcounter{figure}\label{fig:02} Figure \arabic{figure}.}{Exit probability (left panels) and exit time (right panels) as a function of the initial probability $p$ of spin $+1$ for $q=2$ (upper panels) and $q=4$ (bottom panels) for $\beta=0$ (regular graph). The system size $N=100$ and results were averaged over $10^3$ samples.}
\end{figure}

\begin{figure}
\begin{center}
\includegraphics[width=14cm]{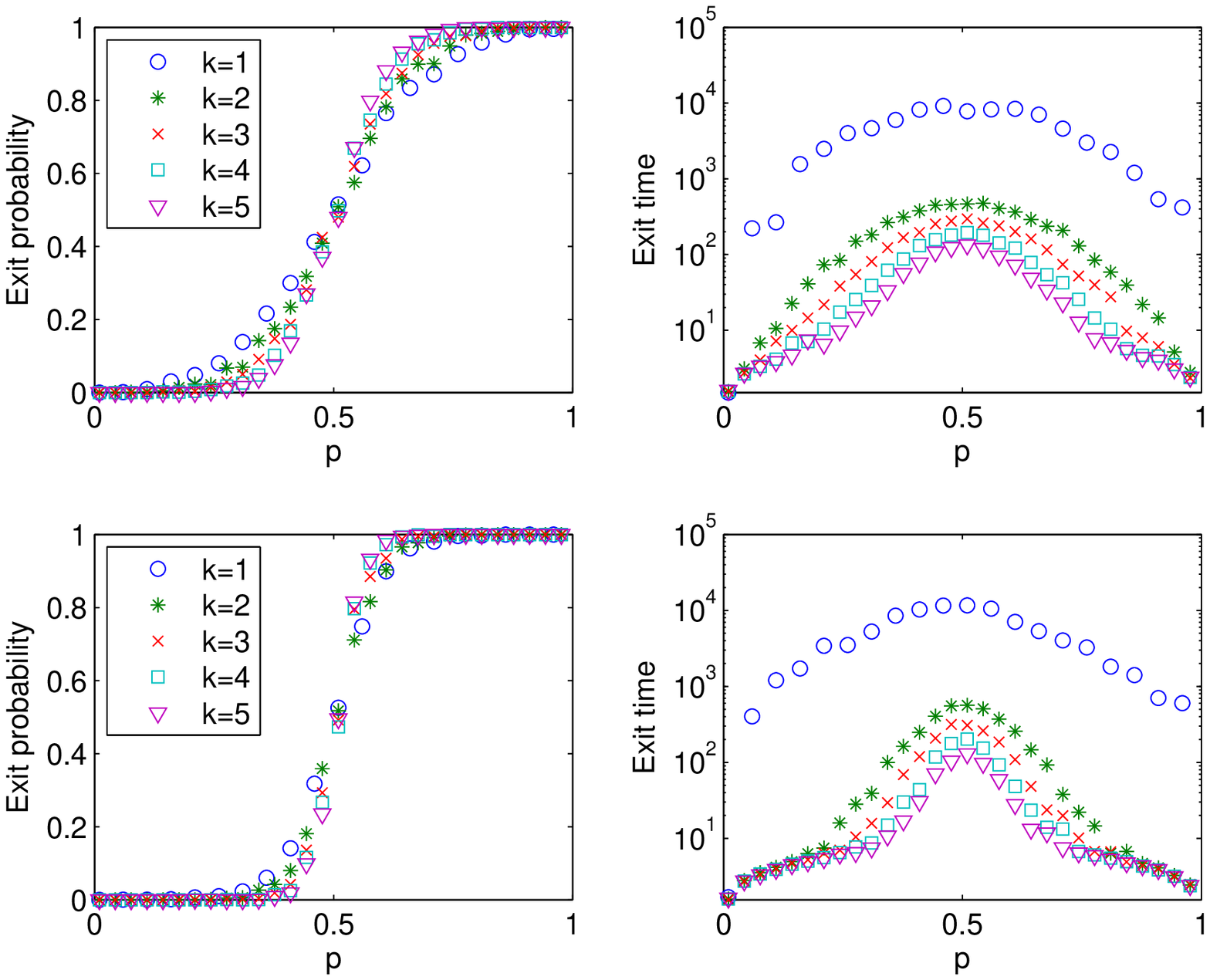}
\end{center}
 \textbf{\refstepcounter{figure}\label{fig:03} Figure \arabic{figure}.}{Exit probability (left panels) and exit time (right panels) as a function of the initial probability $p$ of spin $+1$ for $q=2$ (upper panels) and $q=4$ (bottom panels) for $\beta=0.01$ (small-world). The system size $N=100$ and results were averaged over $10^3$ samples.}
\end{figure}

\begin{figure}
\begin{center}
\includegraphics[width=14cm]{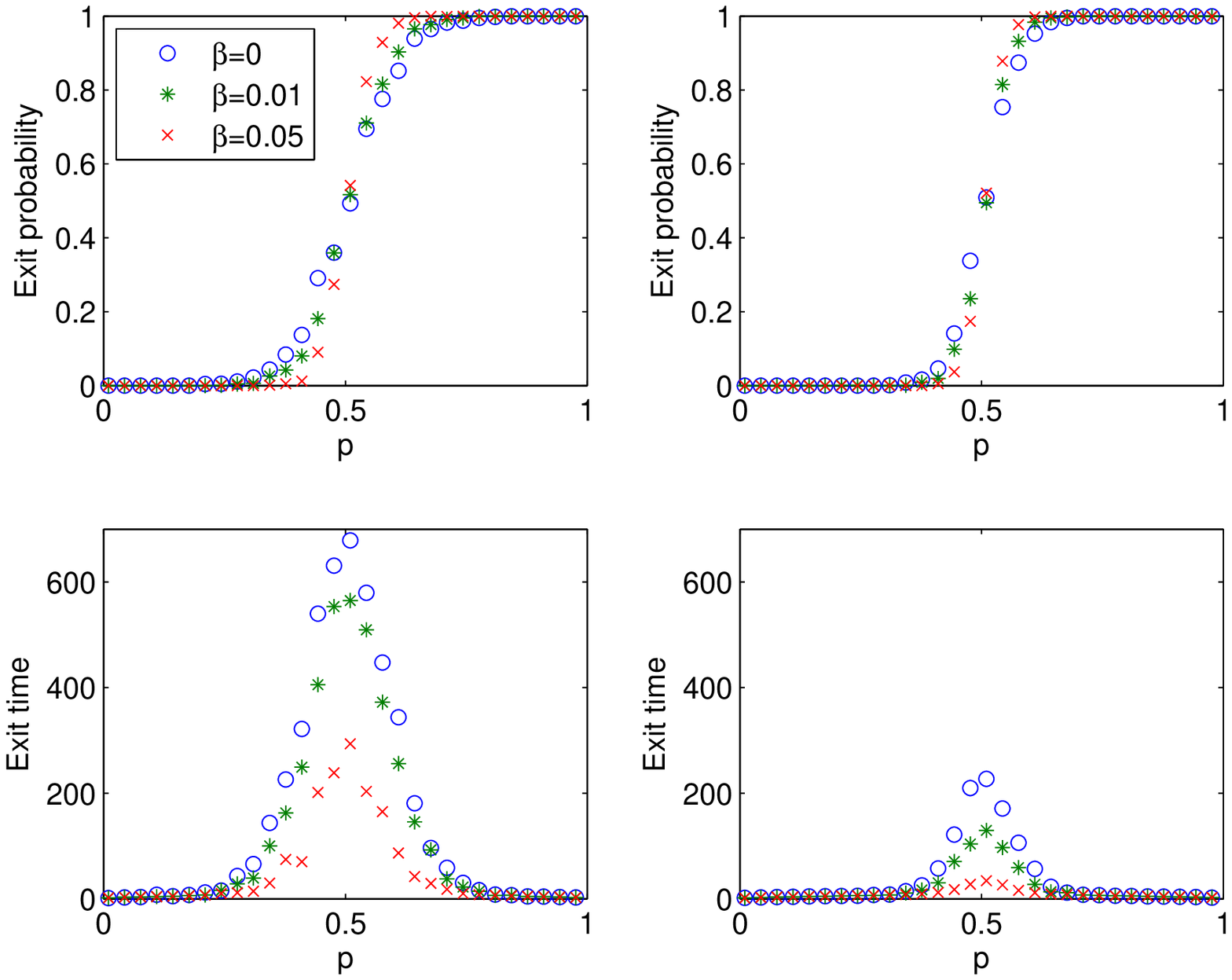}%
\end{center}
 \textbf{\refstepcounter{figure}\label{fig:04} Figure \arabic{figure}.}{Exit probabilities (upper panels) and exit times (bottom panels) as a function of the initial probability $p$ of spin $+1$ for $q=4$ and several values $\beta$. Right panels correspond to $k=2$ and left panels to $k=5$. The system size $N=100$ and results were averaged over $10^3$ samples. The steepness of the exit probability slope increases not only with $k$ but also with the randomness of the graph represented by $\beta$.}
\end{figure}

Now we are ready to examine the role of the second parameter, which describes the level of randomness i.e. $\beta$. In Figs. \ref{fig:02} and \ref{fig:03} we have presented results for two values of $q$ for $\beta=0$ and  $\beta=0.01$. As we see for both values of $q$ the steepness of the exit probability slope increases with $\beta$ (see also Figure \ref{fig:04}). Interestingly, the exit time for $\beta=0.01$ behaves qualitatively different than for $\beta=0$ (compare Figs. \ref{fig:02} and \ref{fig:03}). In both cases the exit time decreases with $k$, but for $\beta=0.01$ (and $\beta=0.05$) there is a dramatic difference between exit time for $k=1$ and other values of $k$. Results are presented in semilog-scale to allow for comparison between different values of $k$. Again results can be fitted using analytical formula (\ref{eq:anal_f}), where $f=f(q,k,\beta)$ is not so easy to determine for arbitrary values of parameters $q,k,\beta$, but for $\beta=0$ can be approximated by $f=\frac{k}{2}+(q-\frac{1}{2})$. It should be remember that our findings are only phenomenologically-based and 
derivation (even approximate) of the formula for the exit probability in a case of arbitrary $k$ and $\beta$ is a challenge for the future. 

\section{Discussion}
We have investigated a special case of a $q$-voter model (with $\epsilon=0$) on a Watts-Strogatz network described by parameters $k$ and $\beta$. We have shown that the exit probability is the S-shaped function for arbitrary value of parameters $q,k$ and $\beta$ and can be fitted by an analytical formula (\ref{eq:anal_f}). Deriving analytical formula (\ref{eq:anal_f}) is a challenge for the future. It should be recall that even for $\beta=0, k=1$ and $q=2$ only approximate calculations are available, although surprisingly four independent approaches \cite{Lam:Red:2008,Sla:Szn:Prz:08,Cas:Pas:11,Gal:Mar:11} give exactly the same result (\ref{Ep2}) that is in a perfect agreement with Monte Carlo results \cite{Prz:Szn:Tab:11} and new results for large lattices obtained by \cite{Tim:Pra:13}. It should be also stressed that there are some contradictory simulation results for $q>2$. It has been argued by \cite{Prz:Szn:Tab:11} that exit probability on one-dimensional lattice does not depend on the system size and can be properly described by Eq. (\ref{eq:Epq}). However, \cite{Tim:Pra:13} have recently shown that for $q>2$ results depend on the lattice size and for large lattices deviate from (\ref{eq:Epq}). Because, they have used a different simulation algorithm that allow to simulate large lattices, additional research is needed in our opinion to clarify the contradiction.

\section*{Acknowledgments}
This work was supported by funds from the National Science Centre (NCN) through grant no. 2011/01/B/ST3/00727.


\section*{References}

\begin{thebibliography}{10}
\bibitem{Cas:For:Lor:09}
C. Castellano, S. Fortunato and V. Loreto, Rev. Mod. Phys. 81, 591-646 (2009)
\bibitem{Cas:Mun:Pas:09}
C.Castellano, M.A.Muñoz, R.Pastor-Satorras, Phys. Rev. E 80, 041129 (2009)
\bibitem{Cas:Pas:11}
C. Castellano, R. Pastor-Satorras,  Phys. Rev. E \textbf{83}, 016113 (2011)
\bibitem{Gal:05}
S. Galam, Europhysics Letters 70, 705–711 (2005)
\bibitem{Gal:12}
S. Galam, \emph{Sociophysics: A Physicist's Modeling of Psycho-political Phenomena}, New York Springer (2012)
\bibitem{Gal:Mar:11}
S. Galam and A. C. R. Martins, Europhysics Letters 95, 48005 (2011)
\bibitem{Kra:Red:Ben:10}
P.L. Krapivsky, S. Redner, E. Ben-Naim, \emph{A Kinetic View of Statistical Physics}, Cambridge University Press (2010)
\bibitem{Lam:Red:2008}
R. Lambiotte and S. Redner, Europhys. Lett. \textbf{82}, 18007 (2008)
\bibitem{Mor:Liu:Cas:Pas:13}
P. Moretti, S. Liu, C. Castellano, R. Pastor-Satorras, J Stat Phys 151, 113-130 (2013)
\bibitem{Nyc:Szn:Cis12}
P. Nyczka, K. Sznajd-Weron, J. Cislo, Phys. Rev. E 86, 011105 (2012)
\bibitem{Prz:Szn:Tab:11}
P. Przybyla, K. Sznajd-Weron, M. Tabiszewski, Phys. Rev. E 84, 031117 (2011)
\bibitem{Sla:Szn:Prz:08}
F. Slanina, K. Sznajd-Weron, P. Przybyla, EPL 82, 18006 (2008) 
\bibitem{Tim:Pra:13}
A. M. Timpanaro and C. P. C. Prado, arXiv:1312.2269v1 (2013)
\bibitem{Wat:Str:98}
D. J. Watts and S.H. Strogatz, Nature 393, 440 (1998)
\end{thebibliography}
\end{document}